\documentstyle[12pt]{article}

\def\beq{\begin{equation}}
\def\eeq{\end{equation}}
\def\beqar{\begin{eqnarray}}
\def\eeqar{\end{eqnarray}}

\def\he#1{\hbox{${}^{#1}$He}}
\def\li#1{\hbox{${}^{#1}$Li}}

\def\yp{\hbox{$Y_{\rm p}$}}

\def\nnu{\hbox{$N_\nu$}}

\def\la{\mathrel{\mathpalette\fun <}}
\def\ga{\mathrel{\mathpalette\fun >}}
\def\fun#1#2{\lower3.6pt\vbox{\baselineskip0pt\lineskip.9pt
  \ialign{$\mathsurround=0pt#1\hfil##\hfil$\crcr#2\crcr\sim\crcr}}}

\begin{document}
\pagestyle{plain}
\thispagestyle{empty}

\rightline{astro-ph/9603009}
\rightline{CERN-TH/96-59}
\rightline{UMN-TH-1424/96}
\rightline{February 1996}

\bigskip

\begin{center}

{\bf Model Independent Predictions of Big Bang
 Nucleosynthesis from \he4 and \li7: Consistency and Implications}

\bigskip

Brian D. Fields$^1$, Kimmo Kainulainen$^2$, Keith A. Olive$^3$, \\
and David Thomas$^4$

\bigskip

{\it $^1$Department of Physics, University of Notre Dame \\
Notre Dame, IN, 46556,  USA}

\medskip

{\it $^2$CERN, CH-1211, Geneva 23, Switzerland}

\medskip

{\it $^3$School of Physics and Astronomy, University of Minnesota \\
Minneapolis, MN 55455, USA}

\medskip

{\it $^4$Department of Physics, University of Florida \\
Gainesville, FL 32611, USA}

\bigskip
\bigskip

{\bf ABSTRACT}

\end{center}

We examine in detail how BBN theory is constrained, and what
predictions it can make, when using only the most model-independent
observational constraints.  We avoid the uncertainties
and model-dependencies that necessarily arise when solar neighborhood D
and \he3 abundances are used to infer primordial D and \he3 via
chemical and stellar evolution models.  Instead, we use \he4 and \li7,
thoroughly examining the effects of possible systematic
errors in each.  Via a likelihood analysis, we find near perfect
agreement between
BBN theory and the most model-independent data.
Given this agreement, we then {\it assume} the correctness of BBN
to set limits on the single parameter of standard BBN, the baryon-to-photon
ratio, and to predict
the primordial D and \he3 abundances.  We also repeat our analysis
including recent measurements of D/H from quasar absorption systems
and find that the near perfect agreement between theory and observation of
the three isotopes, D, \he4 and \li7 is maintained. These results have
strong implications for the chemical and stellar evolution 
of the light elements, in particular for \he3. In addition, our results 
(especially if the D/H measurements are confirmed) have implications
for the stellar depletion of \li7.
Finally, we set limits on the number \nnu\
of neutrino flavors, using an analysis which carefully and
systematically includes all available experimental constraints.
The value \nnu = 3.0 fits best with BBN and a 95\% CL upper limit of 
\nnu $\la 4$
is established.

\newpage

\section{Introduction}

Because of the central role big bang nucleosynthesis (BBN) 
plays in the standard cosmology, it is crucial to understand how
robust the BBN results are.  Consequently there has recently been 
intense scrutiny of different possible sources of uncertainty
in the BBN analysis.  There has been attention to refining
the theoretical calculations, particularly the \he4 yields 
\cite{wssok,he4yields},
as well as quantifying their
uncertainties due to the errors in their input parameters
\cite{kr,skm,kk1,kk2,hata1},
and there has been examination of the observational procedures
and results for possible systematic errors
\cite{osa,csta}.  
Finally, there has been a focus on the model-dependence that
arises in making the link between the observables
and the theory predictions; in this paper we explore in detail
an analysis\cite{fo} that minimizes this model-dependence. 

It is useful to situate our present concerns in
the context of the subject of BBN analysis, which consists of three 
conceptually independent features.
(1) The theory itself rests upon first-principles calculations
and/or detailed laboratory results; the only really free parameter to
the theory calculation is the baryon-to-photon ratio 
$\eta \equiv n_{\rm B}/n_\gamma$, and the output for each value of
$\eta$ consists of the primordial abundances of 
the light elements D, \he3, \he4, and \li7.
(We will consider throughout only homogeneous BBN, and
furthermore we will first study the ``standard'' model 
with $\nnu=3$; later we will relax the latter assumption and look at
possible constraints on \nnu.)
(2) The observations pertinent to BBN are determinations of
the light element abundances in various astrophysical settings;
for the most part, these environments are at
low or zero redshift,
i.e., contemporary to our own, with the exception of 
QSO absorption line system (QSOALS) D/H measurements
which we discuss in more detail below.
As the observable abundances are contemporary, one must invoke
(3) the final facet of BBN analysis, namely some account of galactic
chemical evolution that models the evolution of 
elemental and isotopic abundances from their
primordial state to the present.
In this paper we emphasize that the chemical evolution 
results needed for each of the light isotopes
have very different degrees of model-dependence; 
consequently we will retain only the results which are
the most model-independent, and pursue their implications,
both for BBN and for chemical evolution.

It is necessary to distinguish between the general implications 
chemical evolution gives and its detailed results.
Generally, metal abundances 
(i.e., abundances of elements with $Z \ge 6$)
increase with time, while the
specific histories require
the detailed, quantitative modeling of abundance evolution.
Turning to the light elements, one finds that to infer the primordial 
D and \he3 from local and recent abundances (i.e., solar system
and the ISM) requires detailed modeling.
Of the two, D has the fewest complications
and depends only on the chemical evolution model
features and not at all on stellar evolutionary data, since D/H
is entirely destroyed in the pre-main sequence phases of stars.
\he3 is subject to considerable uncertainties regarding not only
chemical evolutionary parameters but also its stellar processing; the 
possible production in low mass stars and the possible
destruction through mixing effects in the giant branch stars.  
On the other hand, to infer primordial abundances
for \he4 and \li7 requires no detailed modeling,
not because their evolution is simpler, but because we
may determine their evolution empirically.
Specifically, we observe \he4 and \li7 in systems with 
different, and in particular low metallicities.  Since metals
increase with time, we may trace the evolution and infer 
the primordial abundances by extrapolating 
the observed trends to zero metallicity.
This relatively simple procedure is, however, complicated by the very
likely presence of systematic errors in the data, 
and so we will address in detail the
possible instances of and distributions of systematic errors.

Given that the inferred primordial D and \he3 abundances suffer 
from large chemical evolution uncertainties, there are several approaches
one may take when trying to constrain BBN.  The traditional
strategy has been to use the chemical evolution results for
primordial D and \he3, and either trying to quantify the concomitant
uncertainty this introduces \cite{st92,vop,bdf}, or to use a generalized form
of chemical evolution which is designed to be ``generic'' 
\cite{ytsso,wssok,st95}.
This approach has the particular appeal that, as it uses
all of the light elements, it strongly overconstrains the 
one-parameter BBN theory.  On the other hand, 
there remains a concern about the chemical evolution framework
itself, which plays a central role.  As no first-principles calculations
exist for this, it is hard to be certain that the idealizations 
in the available models can capture all 
relevant aspects of the (unavailable) full solution.  This procedure
breaks down when the data begin to be refined enough to
show inconsistencies between theory and observation and the conclusions
for BBN become overly sensitive to highly uncertain modeling. 

An alternative means of addressing the chemical evolution uncertainties
is to minimize the reliance on chemical evolution by using only
its most general features \cite{fo}.
That is, one uses only the inferred primordial \he4 and \li7 abundances,
rather than all four light element isotopes.
Of course, one always has to use chemical evolution at some level,
but for \he4 and \li7 we need only assume that metal abundances
(specifically C, N, O, and Fe) increase with time.
With only this assumption, and a prescription for 
including possible systematic errors,
one may constrain BBN. Note that in the case of \li7,
which is observed in old low metallicity
stellar populations, we also include an uncertainty 
on the degree to which these 
stars did not deplete their \li7 abundance.  

In this paper, we will first review the observational status of all four light
element isotopes.  In the cases of D and \he3, 
we will also review briefly the
arguments which limit their implementation in an analysis of BBN.
Thus we asses
the validity of BBN  using the most model-independent data. We quantify the
success of BBN using likelihood analysis, including theoretical uncertainties
as well as the observational statistical and systematic errors. 
This analysis is discussed in detail, and attention is given
to the significance of the amplitudes of the likelihood functions
as indicators of the goodness of fit.
With this analysis we find, using only \he4 and \li7, that 
BBN theory is consistent with the observations, 
and we determine the best values for $\eta$. We also attempt to quantify the 
reliability of these predictions. 

Though we are privileging \he4 and \li7, this is not to say that
these abundances are known with absolute certainty.  We take as a
starting point the abundances directly from the observations.  
These observations are in remarkable agreement for a
baryon-to-photon ratio which is rather low, $\eta \simeq 1.8 \times
10^{-10}$. A similar range has been also suggested in  \cite{Dar}.
We will also show the effect of modifying the primordial
abundances.  For example, it is sometimes argued that the lithium
abundance observed in halo stars is not the primordial value but
represents a depleted primordial abundance \cite{ddp,ddk}.  We will
show how our results change when the effects of \li7 depletion are
included.

As we have noted, the traditional use of solar and ISM data on D/H 
requires the application of chemical evolution before one may 
use it to constrain BBN; however, there is a real 
chance that the recent measurements of D/H
in quasar absorption systems \cite{quas1,rh}
are in fact true measures of the primordial
D/H abundance.  Therefore, we will also  
show results of the likelihood analysis
which includes \he4, \li7 and D.
In fact we find that the agreement is remarkably good, suggesting 
perhaps that
the putative D in quasar absorption line systems 
is real and very close to primordial levels.

Having determined the success of BBN, we may then use it as has long
been done, as a powerful tool in particle astrophysics.  
On the astrophysics side, we predict the abundances of primordial D and
\he3 on the basis of the \he4 and \li7 data.  These results are
(in addition to being available for testing against quasar absorption line
systems) then useful as inputs in chemical evolution models.
We also determine the potential for BBN to set limits to \li7 depletion; 
these constraints become considerably stronger, when D is included in the BBN
analysis. On the particle side, we let \nnu\ vary, and quantify the allowed 
extra relativistic degrees of freedom. As we will show, if we were
to use BBN, to ``predict" the value of \nnu\, we would find, again remarkably,
that the best value is \nnu\ = 3.0, independent of whether or not we include 
D in the analyses. To wit, the uncertainty in this ``prediction"
depends primarily on the uncertainty in the \he4 observations, which at the
the 2 $\sigma$ level translates to a BBN upper limit of about 4 on \nnu.

The paper is organized as follows:  
The observational data on each of the four light element
isotopes is discussed in section 2, as is the role of chemical evolution in 
deriving primordial abundances of each isotope. 
We will carefully describe our choices
of the input abundances and their statistical and systematic uncertainties.
In section 3, we describe our likelihood analysis method
and present results for BBN consistency in section 4.
Included in section 4, is our discussion of the potential for constraining 
\li7 depletion in stars, and our analyses including D with an assumed 
primordial value taken from the QSOALS measurements \cite{rh}.
In section 5, we discuss the implication of these results on particle 
astrophysics constraints, particularly, the value of \nnu\ .
Finally, in section 6, we draw our final conclusions.

\section{Data}

The literature on light element abundance measurements is
large, and is well-reviewed in, e.g.,
references \cite{wssok,rw,elba}.
Here we will give only a brief summary of the 
current best abundances for each light isotope,
with emphasis on new developments,
particularly the possible detection of D/H in quasar
absorption line systems (QSOALS).  We will also 
discuss the issue of systematic errors in the data, 
and the possible forms such errors might take.

\subsection{Deuterium}
As discussed in \S 1, until recently D abundances were only available for 
the solar system and the local ISM.  The solar value for D/H = (2.6 $\pm 0.6 
\pm 1.4) \times 10^{-5}$ (see \cite{scostv} for a recent discussion) has
a considerable uncertainty. Individual ISM observations are much more
accurate \cite{linetal} but they appear to show dispersion along different 
lines of sight through the ISM \cite{fer,linsky}. This result is unexpected in 
standard models of chemical evolution, and again suggests their incompleteness. 
Moreover, neither of these measurements is 
directly attributable to a primordial 
value.  The D (and \he3) in these systems has suffered a considerable 
degree of processing, and so determining a primordial abundance
from these requires detailed chemical evolution modeling.
As we have argued, our approach is to avoid such modeling,
and so we will not employ these results.

The observational situation for D is rapidly changing
however, and is sparking a renewed interest in D evolution.  
Observers have employed high-resolution spectra of
quasar absorption line systems to determine 
the D/H abundance via its isotope-shifted Lyman-$\alpha$
line.  Initial reports \cite{quas1} gave a high abundance,
D/H $\sim 2 \times 10^{-4}$,
but there remained a significant chance of D being
mimicked by an interloping H cloud, which would lead
to an overestimate of the D abundance.  Recently,
new observations \cite{rh} of this same system have 
in fact resolved this original system into two components
each with an abundance comparable to the old result. 
The weighted average of these two measurements is 
\cite{rh},
\beq
\left(\frac{\rm D}{\rm H}\right)_{z=3.32} = (1.9 \pm 0.4)\times 10^{-4} \ \ 
\label{eq:d}
\eeq
The resolution into two components dramatically
reduces the likelihood of an interloper, strongly suggesting 
that the observations are really measuring D.
Caution is still warranted however, (as it is with any new observation)
until the systematic uncertainty in these observations is clarified and 
more importantly until these results can be confirmed in the QSOALS in other
directions.  For these reasons, we do not include the 
D/H results in our first BBN analysis. We do however, repeat our analyses
using the primordial value for D/H from Eq. (\ref{eq:d}).

\subsection{Helium--3}
Abundances of \he3 are only available for the solar system and the
ISM. The solar value for \he3/H is (1.5 $\pm 0.2 \pm 0.3) \times 10^{-5}$
and varies between (1 to 5) $\times 10^{-5}$ in galactic
 HII regions \cite{bbbrw} (see also \cite{scostv} for a general discussion).
Thus as in the case of D, one must use detailed chemical evolution 
models to extrapolate a primordial abundance.
This procedure is particularly uncertain at the moment
since it is even unclear, on the basis of stellar evolution
theory, whether one expects \he3 to increase or
decrease with time.  On the one hand, \he3 has long been thought
to be produced by low mass stars \cite{rst,it}, an idea apparently
supported by the observation of high \he3 abundances
in  several planetary nebulae \cite{rood}.
On the other hand, chemical evolution models
including such \he3 production lead to overproduction
of \he3 relative to the observations 
\cite{orstv,galli,scostv,dst}.  

There has been some recent progress on this question as
the now venerable models for \he3 production 
have been updated \cite{vw,wwd}.
The different groups agree in finding that there
is net production in the standard model.
It has been suggested
\cite{hogan,ch,wbs,sb}, however, that in fact stellar mixing might destroy \he3.
The inclusion of these effects certainly goes a long way in alleviating the 
problems caused by a high primordial D/H abundance \cite{scostv,bm,osst}.
Indeed, a thorough study of \he3 evolution along with other
isotopes produced by low-mass stars would help to
clarify the situation.

Given the compounded uncertainty regarding the evolutionary history of
\he3 from both stellar and chemical evolution,
we know of no model-independent way to obtain a
statistically significant primordial abundance for \he3 and therefore 
we do not include it in our analysis.
We emphasize, however, that \he3 remains an important tool for chemical
evolutionary models and perhaps also for stellar evolutionary models as
the destruction of \he3 has important consequences on the abundances of 
other isotopes.

\subsection{Helium--4}
The most useful site for obtaining \he4 abundances 
has proven to be H {\sc ii} regions in irregular galaxies.
Such regions have low (and varying) metallicities, and
thus are presumably more primitive than such regions in
our own Galaxy.  Because the \he4 abundance is known for regions of
different metallicity one can trace its evolution as a function 
of metal content, and by extrapolating to zero metallicity
we can estimate the primordial abundance.

There is a considerable amount of data on \he4, O/H, and N/H
in low metallicity extragalactic HII regions \cite{p,evan,iz}.
In fact, there are over 50 such regions observed with metallicities ranging 
from about 2--30\% of solar metallicity. The data for \he4 vs.\ either
O/H or N/H is certainly well correlated and an extensive analysis \cite{osa}
has shown this correlation to be consistent with a linear relation.
While individual determinations of the \he4 mass fraction $Y$
have a fairly large uncertainty ($\Delta Y \ga 0.010$), the large
number of observations 
lead to a {\it statistical} uncertainty
that is in fact quite small.  A recent calculation \cite{osa,osc}
gives
\beq
\label{eq:he4}
Y_p = 0.234 \pm 0.003 \rm (stat) \pm 0.005 (syst)
\eeq
with similar central values and statistical errors 
obtained by other groups.

We have claimed that the primordial abundance of \he4 used below
is devoid of any serious dependence on galactic chemical evolution. The role
chemical evolution plays here is simply to verify that the linear extrapolation 
to zero metallicity is a meaningful estimate of the primordial abundance \yp.
The basic assumption that metal abundances increase linearly with time has
already proven to be a secure ansatz; 
in addition to the strong correlation of the 
data between different metallicities, we are further assured of the accuracy 
of the estimate because the source of the data is derived 
from very low metallicity environments.

Despite the very small
statistical uncertainty in the fit shown in Eq.\ (\ref{eq:he4}), 
more serious complications arise
in the process of extracting abundances from
linestrengths.  As pointed out elsewhere \cite{csta}, assumptions
about the H {\sc ii} region and model-dependencies 
could introduce a significant systematic error.   While the value
of $\sigma_{\rm sys} = 0.005$ we quote above attempts to estimate
this error (see e.g.\ \cite{skill,osa}), 
its magnitude and distribution are not well understood.
Thus in our analysis, we will examine the effect of different
assumptions concerning the systematic errors.

\subsection{Lithium--7}
The primordial \li7 abundance is best determined by studies of
the Li content in various stars as a function of metallicity
(in practice, the Fe abundance).  At near solar metallicity, the Li abundance
in stars decreases with decreasing metallicity,
dropping to a level an order of magnitude lower
in extremely metal poor Population II halo
stars  with [Fe/H] $\la -1.3$ ([Fe/H is defined to be the 
$\log_{10}$ of the ratio of Fe/H relative to the solar value for Fe/H).  
At lower values of [Fe/H], the Pop II abundance remains constant 
down to the lowest metallicities measured, (some with [Fe/H] $< -3$ !)
and form the so-called ``Spite plateau.''
With Li measured for nearly 100 such stars,
the plateau value is well established.   We use the recent results
of \cite{mol} to obtain the \li7 abundance in the plateau
\beq
\label{eq:li7}
\frac{\li7}{\rm H} = (1.6 \pm 0.1) \times 10^{-10}
\eeq
where the errors are statistical only.  Again, if we employ the basic 
chemical evolution conclusion that metals increase linearly with time,
we may infer this value to be indicative of the primordial Li abundance.

One should be aware that there  are considerable systematic uncertainties
in the plateau abundance.  First there are uncertainties that arise even if 
one assumes that the present Li abundance in these stars is
a faithful indication of their initial abundance. The actual \li7 
abundance is dependent on the method of deriving stellar
parameters such as temperature and surface gravity, and 
so a systematic error arises due to uncertainties in
stellar atmosphere models needed to determine abundances.
While some observers try to estimate these uncertainties, 
this is not uniformly the practice.  
To include the effect of these systematics,  we
will introduce the asymmetric error range $\Delta_1 = {}^{+0.4}_{-0.3}$
which covers the range of central values for \li7/H, when different methods
of data reduction are used (see eg. \cite{osc}).

Another source of systematic error in the \li7 abundance arises due to 
uncertainty as to whether the Pop II stars actually {\it have} preserved all 
of their Li.  It has been suggested that, e.g., rotational effects could 
reduce the initial Li abundance to a much lower level, and models have been 
advanced which claim to do so while maintaining the plateau behavior with
respect to metallicity and temperature \cite{ddp,ddk}.  
While the detection of the more fragile isotope \li6 in two of
these stars may argue against a strong depletion \cite{sfosw},
it is difficult to exclude depletion of the order of a factor of two.
Furthermore there is the possibility that the primordial Li 
has been supplemented, by the time of the Pop II star's 
birth, by a non-primordial component arising from cosmic ray interactions 
in the early Galaxy \cite{wssof,fos,vcfo}.  While such a contribution cannot 
dominate, it could be at the level of tens of percent \cite{wssof}.  Note
that this effect acts only to correct {\it downwards} the true
primordial Li abundance from the plateau level.
To allow for these possibilities, we will investigate the effect of a second 
Li systematic error, having the range $\Delta_2 = {}^{+1.6}_{-0.3}$.
In fact, in \S \ref{sec:dep}, we will turn the problem around, and by using 
the D measurements in QSOALS we will offer BBN {\it constraints}
on the level of Li depletion.

\section{Likelihood Analysis}

Monte Carlo and likelihood analyses have proven to be useful tools
in testing the consistency of BBN \cite{kr,skm,kk1,kk2,hata1,fo,hata2}. 
As the observational
determinations of the light element abundances improve, there is little
justification for neglecting the uncertainties in the BBN predictions.
For \he4, these uncertainties are dominated by the uncertainty in the
neutron half-life which have been dramatically reduced in the last
several years, and lead to deviations (1$\sigma$) $\la 0.001$.
For D and \he3, the uncertainty in the BBN yields are less than 10\%.
Of the four light element isotopes, the largest uncertainty resides with
\li7, where the 1 $\sigma$ deviations remain as large as 20--25\% 
\cite{hata1}.

If we restrict our attention to the standard big bang model, in the
context of the standard electroweak model with three neutrino flavors,
there is one single unknown parameter in the standard model of big
bang nucleosynthesis, the baryon-to-photon ratio, $\eta$.  For a given
value of $\eta$, the uncertainties in the BBN calculation of the light
element abundances stem from the uncertainties in the nuclear (and
weak) interaction rates employed. When we consider deviations from the
standard electroweak theory, which can sometimes be parameterized by
changing the number of light neutrino degrees of freedom, the
abundances will be sensitive to the choice of $N_\nu$.  As we stated
earlier, it will be sufficient to use only two of the light element
abundances to test this one-parameter theory (one is enough to constrain
it).  If and when they are applicable, the others will even more strongly 
test/constrain the theory.

The Monte Carlo calculations in BBN make available a distribution of
abundances at each value of $\eta$, based on the uncertainties in the
nuclear and weak interaction rates, which we take to be Gaussian distributed. 
As in \cite{fo}, we will use as the starting point the Monte Carlo results from 
Hata et al.\ \cite{hata1}.  Thus for each of the light elements we obtain a
theoretical distribution function, which depends on $\eta$ and on
the element abundance.  For example, we can begin with a likelihood
distribution from the BBN calculation for \he4:
\beq
L_{\rm BBN}(Y,\eta) 
  = {1 \over \sqrt{2 \pi} \sigma_1} e^{-\left(Y-Y_{\rm
BBN}\left(\eta\right)\right)^2/2\sigma_1^2}
\label{bbn}
\eeq
where $Y_{\rm BBN}(\eta)$ is the central value for the \he4 mass
fraction produced in the big bang, and $\sigma_1(\eta)$ is the
uncertainty in that value derived from the Monte Carlo
calculations. Note that there is a mild dependence on $\eta$ in
$\sigma_1$.

There is also a likelihood distribution based on the observations.  In
this case we have two sources of errors as discussed above, a
statistical uncertainty, $\sigma_2$ and a systematic uncertainty,
$\sigma_{\rm sys}$.  Unfortunately, there is no well defined way to
treat the systematic errors.  One possibility is to assume that the
systematic error is described by a top hat distribution
\cite{hata1,osb}.
In this case, the convolution of the top hat
distribution and the Gaussian (to describe the statistical errors in
the observations) results in the difference of two error functions
\beqar
L_{\rm O}(Y,Y_{\rm O}) =  \frac{1}{2(\sigma_{\rm sys+}+\sigma_{\rm sys-})}
  & & \left[
     {\rm erf}\left({Y - Y_{\rm O} + \sigma_{\rm sys-}  
               \over \sqrt{2} \sigma_2}\right)   
      \right. \nonumber \\
  &-& \left. 
     {\rm erf}\left({Y - Y_{\rm O} - \sigma_{\rm sys+} 
               \over \sqrt{2} \sigma_2}\right) 
      \right]
\label{erf}
\eeqar
where in this case, $Y_{\rm O}$ is the observationally determined
value for the primordial \he4 mass fraction and we have allowed for the 
possibility of asymmetric systematic uncertainty.  The distribution 
(\ref{erf}) is normalized to one with respect to integration over $Y$.

In addition to the top-hat distribution for the systematic
uncertainty, we have also derived the likelihood functions assuming
that the systematic errors are Gaussian distributed. In this case, the
convolution also leads to a Gaussian, with an error $\sigma^2 =
\sigma_2^2 + \sigma_{\rm sys}^2$.  
Finally, as a third case, we have
also simply shifted the mean value $Y_{\rm O}$ by an amount $\pm
\sigma_{\rm sys}$. In this case $L_{\rm O}$ is also a Gaussian, with
spread $\sigma_2$. 
These functions were similarly derived for \li7.

For \he4, we constructed a total likelihood function for each value of
$\eta_{10} \equiv 10^{10} \eta$, convolving for each the theoretical
and observational distributions
\beq
L^{^4{\rm He}}_{\rm total}(\eta) = 
\int dY L_{\rm BBN}\left(Y,\eta\right) 
L_{\rm O}(Y,Y_{\rm O})
\label{findist}
\eeq
An analogous calculation was performed for \li7.  At this point we 
could use the two likelihood functions, and their product, to calculate
(say) 68\% CL intervals in $\eta$.  This would tell us the values of 
the parameter ($\eta$) for which the theory (BBN) best fits the 
data.  It tells us nothing however about whether that ``best'' fit is 
a good fit.  For example, if the observed \li7 abundance were several
standard deviations below the minimum value in the \li7 vs $\eta$ curve, 
then clearly the data fit the theory very poorly.  The likelihood 
function for \li7 would still however show a peak at a value of $\eta$
near that minimum, and would lead to a 68\% CL interval in $\eta$ 
similar to that obtained for a much higher observed value.
If the only errors we had to concern ourselves with were
Gaussian then we could calculate a $\chi^{2}$ from the likelihood, 
and this would then give us the probability that the theory fits the 
data (or more correctly, that the theory fails to fit the data).  
Unfortunately of course, the errors are not all Gaussian and there is 
no standard technique for calculating the goodness of fit.  In this 
paper we estimate the goodness of fit from the height of likelihood 
functions at their peak as described below.

We begin by defining
a renormalized probability 
distribution in $\eta$ by a simple rescaling,  
 such that
\beq
\int L^{^4{\rm He}}_{\rm total}(\eta) d\eta = 1.
\label{lhe4}
\eeq
Likelihood functions like this, derived also for lithium, form the basis of our 
subsequent analysis.  We should note that, while this procedure in
principle throws away the information on whether the observed lithium
abundance is below the minimum value in the lithium vs $\eta$ curve
(as in our example above)  this is never an issue in practice.


We will see that each of the abundances can individually be reconciled with the
one-parameter theory, each predicting a distinct concordance region in $\eta$.
Because of the independence of the observational distributions folded with the
(dependent) theoretical distributions to give our 
likelihood distributions, these 
``measurements" of $\eta$ are all 
independent of each other. Demanding that two
or more, in our case in particular \he4 and \li7, be fit simultaneously, 
constitutes a test for the theory.  To do this in practice, one
examines the product of the individual likelihoods, ${\cal L} = L^{^4{\rm
He}}_{\rm total}(\eta) L^{^7{\rm Li}}_{\rm total}(\eta)$ which
yields information on the goodness of fit based
on the magnitude of this quantity at the peak of the combined
distribution (if any) and of the spread in the allowed values in
$\eta_{10}$.

Before we move on to discuss specific results, it will be
useful to discuss the issue of goodness of fit, and how it is
to be quantified. In the case of BBN the number of ``measurements'' never 
increases; we will always have only 2 to 4 of them.  So there is no hope of
ever being able to apply the central limit theorem to obtain a well defined mean 
and spread in the predicted range of $\eta$ in the usual sense of finding 
``world averages'' of experimental quantities from ensembles of independent 
experimental results.  On the contrary, in BBN one can always go back and refine 
the errors of the individual ``measurements'', thus getting an improved test of 
the theory that way.  Nevertheless, the statistical likelihood analysis will 
always be hampered by the small number of data points.
Indeed, even if all of the errors are Gaussian, the usual
$\chi^2$ analysis only marginally applies for $N = 2-4$.  However, in our case 
the errors and distributions are in fact far from Gaussian.  Though we can still 
define a quantity $\chi^2 = -2 \ln \cal{L}$, its utility is unclear.
However, the combined likelihood function does
carries certain information about the 
goodness of fit; we will choose to access this by examining the significance 
of the amplitude.

To introduce our approach, we first
consider some simple cases to determine what the magnitude at the
peak of the combined distribution will tell us. For example, suppose
that we we have two normalized Gaussian distributions, 
$L_1(x) = (1/\sqrt{2\pi}\sigma_1) e^{-x^2/2\sigma_1^2}$, 
and 
$L_2(x) = (1/\sqrt{2\pi}\sigma_2) e^{-(x-\mu)^2/2\sigma_2^2}$. 
(Think of these as toy models for $L^{^4{\rm He}}_{\rm total}(\eta)$
and $L^{^7{\rm Li}}_{\rm total}(\eta)$ respectively.)
The product of
these distributions is also a Gaussian
\beq
L_1(x) L_2(x) = (1/2\pi\sigma_1\sigma_2) e^{- \mu^2/2 \sigma^2}
e^{-{\sigma^2 \over 2 \sigma_1^2 \sigma_2^2} (x-{\sigma_1^2 \over \sigma^2}
\mu)^2}
\label{ex}
\eeq
where $\sigma^2 = \sigma_1^2 + \sigma_2^2$.
In the event that the two distributions have the same mean value, $\mu
= 0$, then the peak of our combined distribution has a magnitude
$1/2\pi\sigma_1\sigma_2$. When the distributions are offset, even though
the product is still a Gaussian, the magnitude at the peak is now
suppressed by a factor $e^{- \mu^2/2 \sigma^2}$. Clearly for $\mu \gg
\sigma$, this suppression can be significant and indicates a lack of
goodness of fit.  It would seem appropriate, therefore, to compare
the value of the peak of the combined likelihood distribution
(\ref{findist}) with the quantity $(1/2\pi\sigma_1\sigma_2)$. Despite
the fact that the suppression factor in (\ref{ex}) is exponential,
unless $\mu \gg \sigma$, it will take values in the tens of
percent. For example, if $\sigma_1 \simeq \sigma_2$ and if our
distributions are offset by 1 (2) $\sigma$, the suppression factor is
still relatively mild, 0.78 (0.37). Even for a distribution whose
peaks are separated by 3$\sigma$, i.e. distributions that we would
generally judge as being inconsistent, the suppression is only about
0.1.  Thus when we make the comparison at the peak of our combined
distribution to $(1/2\pi\sigma_1\sigma_2)$, any value significantly
less than 1 will indicate a poor fit.

Thus far in our example, we have only considered the case in which 
the likelihood functions, $L(\eta)$ are Gaussian.
However, in general our
likelihood distributions for \he4 and \li7 will {\em not} be
Gaussians.  In particular, as we will see, in the case of \li7, our
distribution is double peaked.  We can however still make the
comparison to $(1/2\pi\sigma_1\sigma_2)$ if we define the respective
$\sigma$'s as the half-width at half-maximum of a local peak.  This
will be made clear when we consider specific examples below.

In considering the bimodality of the Li likelihood function, it is 
useful to adopt a larger perspective.  Over an extended range of
$\eta$, beyond the canonical $\eta_{10} \sim 1-10$ 
range,\footnote{Indeed, if one is testing BBN itself, one should in
principle consider a large range in $\eta$, but of course experience 
with  BBN calculations, as well as other available bounds on $\Omega_{\rm B}$
and hence on $\eta$, lead one to focus straightaway on the canonical range.}
all of the light element abundances are {\it non-monotonic} functions of $\eta$.  
Thus, including a larger range of $\eta$ in one's analysis would yield multiple 
peaks in each isotope's individual likelihood function.
Of course, the ``new'' peaks would not overlap between different isotopes, 
and the predicted range for $\eta$ as given by the combined likelihood function 
would remain unchanged. Hence it is reasonable to independently normalize each 
well-isolated peak in an isotopic likelihood. This procedure is straightforward
for all of the isotopes other than Li, whose peaks are not isolated and 
can overlap to some extent. Thus we normalize over both of these peaks and 
therefore expect a reduced amplitude in the combined distribution.

\section{Model-Independent Odds on BBN}
\subsection{Standard Cases}
The procedure that we set up in the preceding section would be
straightforward to carry out if it were not for the complication of
the treatment of the systematic uncertainties in the data. We will
therefore present results for various different assumptions regarding
the data.  As described in section 2, we will always make comparisons
with respect to our standard case in which the data is described by
Eqs. (\ref{eq:he4}) and (\ref{eq:li7}), namely $Y_p = 0.234 \pm 0.003 \pm
0.005$ and \li7/H = $(1.6 \pm 0.1 ~^{+.4}_{-.3}) \times 10^{-10}$. The
effects of lithium depletion are neglected at this stage, and the systematic 
errors are described by a top-hat distribution.

With these assumptions, we have calculated the likelihood functions
for \he4 and \li7 \cite{fo} which are shown in figure \ref{fig1}.  The
shapes of these curves are characteristic, with one peak for \he4
(whose abundance rises monotonically with $\eta$), and two for \li7
(whose abundance goes
through a minimum).  In this case (and most others) the minimum
theoretical Li is somewhat below most of the observational values and
so the sides of the minimum are favored, leading to the two peaks;
i.e., for a given observational value of \li7, there are two values
for $\eta$ at which this may be achieved. By glancing at figure
\ref{fig1}, it does not take very much statistical machinery to see
that the BBN predictions for \he4 is consistent with the low-$\eta$
value of the \li7 prediction when compared with the observations.
These distributions are clearly consistent.

The combined likelihood, for fitting both elements simultaneously, is
given by the product of the two functions in figure \ref{fig1}, and is
shown in figure \ref{fig2}. As we discussed in the previous section,
we have scaled the combined distribution by a factor
$2\pi\sigma_4\sigma_7$ where $\sigma_4 = .66$ is the half-width at
half-maximum of $L^{^4{\rm He}}_{\rm total}(\eta)$ and $\sigma_7 =
.38$ is the corresponding quantity for $L^{^7{\rm Li}}_{\rm
total}(\eta)$. In this latter case, the distribution $L^{^7{\rm
Li}}_{\rm total}(\eta)$ is clearly far from Gaussian and the value chosen
for $\sigma_7$ corresponds to the low-$\eta$ peak only. Because the
\li7 distribution is nearly equally divided into two peaks we should
expect completely overlapping distributions to yield a peak value of
only $\sim 0.5$ in the combined distribution\footnote{Note that the
resulting peak value of 0.5 does carry some statistical information.
For example, suppose our \li7 distribution takes a form in which the
peaks were different from one another (e.g. if the left peak were a
factor of 10 smaller than the right peak). While we could argue that we
would expect a combined fit of only 0.1, we should also conclude a low
confidence level for consistency. Thus it is not appropriate to scale
out this factor in the combined distribution function. However,
the \li7 peaks are nearly always about equal. }.  
This is what one
finds in figure \ref{fig2}, which does show concordance, the peak is
indeed close to 0.5 at $\eta_{10} = 1.8$.

The allowed 68\% CL
and 95\% CL ranges are
\begin{eqnarray}
1.6 & < \eta_{10} & < 2.8 \qquad 68\% {\rm CL} \nonumber \\ 1.4 & <
\eta_{10} & < 3.8 \qquad 95\% {\rm CL}
\label{res1}
\end{eqnarray}
Note that these intervals are based on standard statistical techniques,
and are not dependent on our chosen method of assigning a 
goodness-of-fit to the height of the likelihood peak.

Thus, for this ``standard'' case, we find that 
the abundances of 
\he4 and \li7 are consistent, and select an $\eta_{10}$ range which
overlaps with (at the 95\% CL) the longstanding favorite range around 
$\eta_{10} = 3$. Further, by finding concordance (in this
case) using only \he4 and \li7, we infer that if there is problem 
with BBN analysis, it must arise from D and \he3 and is thus tied to chemical 
evolution. The most model-independent conclusion is that standard
BBN  with $N_\nu = 3$ is not in jeopardy, but there are problems 
with our detailed understanding of D and particularly \he3
chemical evolution.

The concordance range for $\eta$ given in Eq.(\ref{res1}) allows us to make 
definite predictions for the primordial abundances of D and \he3.
The 95 (68) \% CL ranges in (\ref{res1}) corresponds to
$5.5 (8.9) < {\rm (D/H)} \times 10^5 < 28 (22)$ with a best value for
D/H = $1.8 \times 10^{-4}$ at $\eta_{10} = 1.8$. 
As we have already noted,
this value for D/H agrees incredibly well with the QSOALS observations
in \cite{rh}. For \he3, we have,
$1.4 (1.7) < {\rm (\he3/H)} \times 10^5 < 2.7 (2.5)$, with a best value
\he3/H = $2.3 \times 10^{-5}$ which, it should be noted, is larger than the
solar value of \he3/H.

In the preceding analysis, we have described the systematic
uncertainty by a top-hat distribution. This prescription gives a 
tight set of errors which provides a stringent test to BBN, however,
it also gives the most restrictive
bounds on $\eta$.  
Thus when we relax this assumption
and treat the systematics as Gaussian distributed, we will expect
concordance over a greater range.  In figure \ref{fig3}, we show the
\he4 and \li7 likelihood functions when the systematic errors are
Gaussian.  It should not be surprising that the concordance is present
at the same level though the widths of the distributions are larger.
In this case, $\sigma_4 = .86$ and $\sigma_7 = .53$.  The combined
distribution shown in figure \ref{fig4} is also similar to that in
figure \ref{fig2}, the peak is again close to 0.5 and the width of the
distribution is larger, placing greater weight at higher $\eta$.  This
distribution leads to a predicted range for $\eta$, $1.4 (1.7) <
\eta_{10} < 4.3 (3.4)$ The peak of the distribution is unchanged at
$\eta_{10} = 1.8$. This range in $\eta$ corresponds to 
$4.6 (6.5) < {\rm D/H} \times 10^5 < 28 (20)$ 
and $1.3 (1.5) < {\rm (\he3/H)} \times 10^5 < 2.7 (2.4)$.

Having found concordance for BBN in the standard model, we now 
vary the parameters of the distributions---in particular, those
for the systematic errors---and examine the results.

\subsection{\li7 Depletion}
\label{sec:dep}

Next we would like to test the uncertainty in \li7 given by
$\Delta_2$.  That is, what is the effect of \li7 depletion and
cosmic-ray production?  We will first assume that the \li7/H abundance
is shifted up by $\Delta_2$, that is, we assume a factor of 2
depletion in the Pop II halo stars.  We will treat the error in
$\Delta_1$, as well as the systematic error in \he4 as Gaussian, this
being a more conservative approach than the top-hat method.

By shifting the observational primordial \li7 abundance upwards, we are in 
effect separating the two peaks in the \li7 likelihood distribution. The
results for this case is shown in figure \ref{fig5}. There is still
some overlap between the \he4 and \li7 distributions. In this case
however, there is a distinction between the two lithium peaks. The
half-width of the low-$\eta$ peak is 0.19 while for the high-$\eta$
peak it is 0.68. In the combined distribution, we have scaled the
overlap regions according to these respective peaks, which can be
justified in this case since the combined distribution now shows two
distinct distributions as seen in figure \ref{fig6}. This combined
likelihood distribution is in clear contrast to the previous cases
considered.  Note the peak value at low-$\eta$ is only about 0.15 (it
is even lower for the high-$\eta$ peak).  We do not consider this to
be a good fit of theory to data; yet on the basis of this result 
relying \he4 and \li7 alone, we cannot quite exclude depletion at 
this level.

To salvage the consistency of BBN with a factor of 2 depletion in Li 
one might consider an increase the systematic uncertainty in
\he4 by a factor of 2, to $\sigma_{\rm sys} = 0.01$. By doing so,
we would flatten and greatly broaden the \he4 likelihood distribution. This 
would of course produce a greater overlap with the high-$\eta$ peak in the \li7
distribution.  However, we can not place any great significance
to the improved overlap in this case, because by increasing the systematic
uncertainty to this extent, we have in effect taken all of the
predictive power away from \he4. ( The \he 4 distribution in this case
is now mildly peaked at $\eta_{10} \simeq 1.8$ at a value of 0.2, and
is only slowly decreasing out past $\eta_{10} = 10$. ) Thus we are
only really considering \li7 in this case, and some level of
consistency is almost trivially guaranteed.

Another possibility to salvage BBN consistency with \li7 depletion
would be to shift up the primordial \he4 abundance.  
However, a shift up by an amount $\sigma_{\rm sys} = 0.005$ to
a primordial value $Y_p = 0.239 \pm 0.003$ results in a terrible fit
in which the combined distribution peaks at a value of 0.05!  Only if
the primordial abundance is shifted by at least 0.01 is the overlap
between \he4 and \li7 significant enough to give a reasonable combined
distribution.  It is our opinion, however, that in this case we are
straying far from the raw observations which show the remarkably good
compatibility in figures 1-4. 

Let us also  comment that if one assumes that perhaps 20\% of
the observed \li7 were produced by cosmic-ray nucleosynthesis rather
than BBN, then compatibility is further improved.  If we shift the
\li7 down by $\Delta_2$, the two \li7 peaks begin to merge and there
is a strong overlap in the distribution functions.  The combined
distribution peaks at $\eta_{10} = 2.0$ in this case.  

Finally, we point out that in order to bring the probability of overlap
of the theoretical and observational values of the Li
abundance\footnote{ 
This is given by an integral equation similar to (7) for lithium
with the lower limit of integration set to the theoretical minimum and 
letting the observational value vary until the value of the integral becomes 
0.05.} 
below the 5 per cent level, would require reducing the observationally 
inferred abundance by an order of magnitude.  This would be the lowest 
observational value BBN could be in agreement with, but at present the
limit does not seem at all interesting. 

\subsection{Primordial D/H}

It is interesting to note that the central (and strongly)  peaked
value of $\eta_{10}$ determined from the combined \he4 and \li7 likelihoods
is at $\eta_{10} = 1.8$.  The corresponding value of D/H is 1.8 $\times 
10^{-4}$, very close to the value  of D/H in quasar absorbers
in the published set of observations
\cite{quas1,rh}.
It is not clear if this is a coincidence or if we really have evidence
that three of the
light element abundances point to the same value of $\eta_{10}$.

Noting the perhaps still preliminary nature of the QSOALS D/H
measurements, we move ahead and perform the likelihood analysis for
the three light elements D, \he4 and \li7.  To include D/H, we
proceed in much the same way as with the other two light elements.  We
compute likelihood functions for the BBN predictions as in
Eq. (\ref{bbn}) and the likelihood function for the observations using
D/H = $(1.9 \pm 0.4) \times 10^{-4}$.  In this case, since the
systematic error is not stated, it is neglected here 
and we adopt a simple Gaussian form for the statistical errors. 
These are
then convolved as in Eq.  (\ref{findist}).  
In figure 7, the resulting normalized
distribution, $L^{{\rm D}}_{\rm total}(\eta)$ is super-imposed on
distributions appearing in figure 1. 
It is indeed startling how the three peaks, for
D, \he4 and \li7 are literally on top of each other.  In figure 8, we
show the combined distribution which has now been scaled by $((2
\pi)^{3/2} \sigma_2 \sigma_4 \sigma_7)^{-1}$ with $\sigma_2 = .29$.
We have a very clean distribution and prediction for $\eta$:
\begin{eqnarray}
1.65 & < \eta_{10} & < 2.06  \qquad 68\% {\rm CL} \nonumber \\
1.50 & < \eta_{10} & < 2.37  \qquad 95\% {\rm CL}
\label{res2}
\end{eqnarray}
with the peak of the distribution at $\eta_{10} = 1.75$.  
The absence of any overlap with the high-$\eta$ peak of the \li7
distribution has considerably lowered the upper limit to $\eta$. 
Overall, the concordance limits in this case are dominated by the 
deuterium likelihood function, which, we should caution again,
is based on an observation along a single line of sight.

A high value of D/H, as measured in the QSOALS, requires a significant
amount of D destruction over the history of the Galaxy. Although this
alone is not necessarily problematic, since chemical evolution models
can be constructed to account for such factors of deuterium destruction
\cite{orstv}, it compounds the problem of \he3 overproduction.
The fact that \he3 was problematic even at higher $\eta_{10}
\sim 3$, was our motivation for neglecting \he3 to begin with.

Finally, we consider the effects of the D/H measurements when
\li7 is assumed to be depleted.  In figures 9 and 10, we
show the individual and combined likelihood functions which can be
compared with those in figures 5 and 6 with D/H.  In the combined
distribution, the high-$\eta$ peak is now gone, and the low-$\eta$
peak is hardly significant.  Recall (\S\ref{sec:dep}) that \li7
depletion could be tolerated by either an enlargement of the \he4 
systematic error or a shift upwards in the \he4 abundance, which would
would allow significant overlap with the high-$\eta$ peak. However, when 
the D/H observations are included in the analysis, the high-$\eta$ peak is
absent in the combined distribution and the low-$\eta$ peak takes a
value of less than 0.05.  Once the D/H measurements are confirmed,
they will provide a strong constraint on the degree of \li7 depletion
in halo stars when the consistency of big bang nucleosynthesis is
assumed.

\section{Constraints on \nnu}

In the previous section we have demonstrated the health of
BBN when model-independent constraints are applied; we arrived
at these conclusions without first assuming the validity of BBN.  
Now we assume its validity and, as has been done traditionally, use it 
to constrain new physics.  In particular, we will consider the effects of
our analyses on the limit to the number of neutrino flavors, \nnu.

The light element isotope which is most sensitive
to the number of neutrino flavors
is \he4.  In fact, it has been common
\cite{wssok} to express (by means of a fit)
the \he4 mass fraction as a function of $\eta_{10}$, \nnu,  and the neutron
meanlife. For example, near $\eta_{10} = 2$ and \nnu\ = 3,
we have, 
\beq
Y = 0.2262 + 0.0131 (N_\nu-3) + 0.0135 \ln \eta_{10} + 2 \times 10^{-4}
(\tau_n -887)
\label{ynnu}
\eeq
Therefore, given an upper limit to the \he4 mass fraction {\em and} a
lower limit to $\eta$ one can derive an upper bound to \nnu.
Indeed, unless the lower bound to $\eta_{10}$ is greater than 0.4
(something we are assured of now by the D/H measurements),
there is no upper limit to \nnu\ \cite{ossty}.
In \cite{ytsso}, it was argued that the solar value of D+\he3 could provide
a reasonable lower bound to $\eta$.  Recall that this argument was made
before the evidence from planetary nebulae pointed towards the 
need for \he3 production in low mass stars and before the QSOALS measurements
of extremely high D/H. 
Therefore, the role of chemical evolution was not believed to be critical.
Upper limits to \nnu\ were then derived by inserting  the 2 $\sigma$ upper
limit to $Y$, the lower bound on $\eta$ from D + \he3, and the lower bound 
on $\tau_n$ into Eq. (\ref{ynnu}).  The result found in \cite{wssok} 
was \nnu\ $< 3.3$.

The methods for deriving \nnu\ have since been refined incorporating a more
statistical meaning to the bound. In \cite{osa}, using the best central values
and their associated uncertainties in an expression similar to (\ref{ynnu})
gave  a best estimate, \nnu\ $= 2.2 \pm 0.3 \pm 0.4$,
(where the statistical error is taken from the observational determination of 
$Y$ and the neutron mean life, and the systematic error from 
$^4$He and from $\eta$,  
assuming that $\eta_{10} = 3.0 \pm 0.3$).
Since one could well imagine theories that would effectively {\em
lower} the value of $N_\nu$ as well as increase it, one has to, in
the broadest sense take the bound
seriously, and accept that it might show  preference for some
extension of the standard model predicting less helium. 
The (2 $\sigma_{\rm stat} + \sigma_{\rm sys}$) upper limit was 
found to be $N_\nu < 3.1$ \cite{osa}. A similar bound, \nnu\ $< 3.04$
was obtained from a Monte Carlo analysis \cite{kk1} and an even stronger
bound was obtained using a more refined treatment of the solar D + \he3 
argument, \nnu\ = $2.0 \pm 0.3$ \cite{hata2}. If we assumed a priori that
\nnu\ $ > 3$, say, then we could use a Bayesian approach to relax
the upper limit to \nnu\ \cite{osb,fko1}. An alternative argument for 
relaxing the constraint based on a stochastic approach to chemical
evolution was made in \cite{cstb}. Also, for a critical 
review on the history of the
bound on \nnu\ see also \cite{Sarkar}.

All of the above methods for placing a bound on \nnu\ rely on \he3 and in 
particular, the D + \he3 argument for a lower bound to $\eta$.
However, as we have emphasized repeatedly in this work, the uncertainties in
the evolutionary history of \he3 make it a poor indicator of \nnu\ or 
of the validity of BBN.
If instead we use the results
found in \S 4, for $\eta$ based on \he4 and \li7, we find as the best estimate
for \nnu,
\beq
N_\nu = 2.99 \pm 0.23 \pm 0.38 {~}^{+ 0.11}_{- 0.57} 
\label{Nlim2}
\eeq
showing no particular preference to $N_\nu < 3$, in fact preferring the 
standard model result of $N_\nu = 3$ and leading to $N_\nu < 3.90$ 
at the 95 \% CL level when adding the errors in quadrature.
(Since the central value in Eq.\ (\ref{Nlim2}) is dramatically
close to 3, we show two decimal places, but of course
the errors imply that the precision of the agreement is
somewhat fortuitous.)
In (\ref{Nlim2}), the first set of errors are the
statistical uncertainties primarily from the observational
determination of $Y$ and the measured error in the neutrino half life
$\tau_n$. The second set of errors is the systematic uncertainty
arising solely from $^4$He, and the last set of errors from the
uncertainty in the value of $\eta$ and is determined by the combined
likelihood functions of \he4 and \li7, ie taken from
Eq.\ (\ref{res1}). We view Eq.\ (\ref{Nlim2}) as a further (and remarkable)
confirmation of standard BBN.

Since we have shown our likelihood results when quasar D/H is included, 
we can also derive the best value for \nnu\ when D, \he4 and \li7 are
all included in the analyses.  In this case,
\beq
N_\nu = 3.02 \pm 0.23 \pm 0.38 {~}^{+ 0.06}_{- 0.18} 
\label{Nlim3}
\eeq
leading again to a 95 \% CL limit of \nnu $< 3.9$.
In Eq. (\ref{Nlim3}), the central value is slightly higher due to the 
central value for $\eta_{10}$ being slightly lower in this case (1.75 rather than
1.80) and the error due to the uncertainty is considerably smaller especially
on the low side (corresponding to the high side on $\eta$).

\section{Conclusions}

We have argued that because D and \he3 are observed only in the solar system 
or the local ISM, determinations of their primordial abundances are
particularly uncertain and dependent on models of galactic chemical 
evolution.  It is therefore important to
determine the status of BBN without them.  Using a likelihood analysis
we find that the compatibility of the
\he4 and \li7 constraints depends on the
assumptions one makes about the size and distribution of the possible
systematic errors. 
Nevertheless, for the most ``standard'' assumptions, i.e.\ taking the data
on \he4 and \li7 at face value, we find that these two isotopes
are quite compatible and
give BBN concordance to a high degree of confidence. The baryon-to-photon ratio
selected by this concordance is
centered around $\eta_{10} \sim 1.8$ and ranges up to 3.8 (or 4.3 if all
errors are assumed to be Gaussian).

While the putative determinations of D in quasar absorption line
systems are still in their infancy, the technique can be a powerful probe.
Upon introducing into our analysis the relatively high D abundance
reported for several systems, we find that the agreement with
\he4 and \li7 is striking.  It is crucial to further observe
and better understand these systems, which can provide a decisive
constraint on BBN.

Having found BBN theory to be consistent with observation, we turn
to its implications.  In the realm of astrophysics, we find that the
needed primordial D and \he3 abundances challenge our understanding of
the chemical evolution of these nuclides, particularly \he3.  We find
that the presence of \li7 depletion does not improve the BBN
agreement, and further, if one adopts the high D abundance suggested
by the quasar measurements, one can determine the primordial \li7
abundance and thus tightly constrain the degree of depletion.  In
the realm of particle physics, we revisit the
\nnu\ counting argument, with a careful treatment of all
the data giving a best value for \nnu\ = 3.0 and a 95\% CL
upper limit of $\nnu \la 3.9$.

\bigskip

We would like to thank Craig Copi, David Schramm, and Jim Truran
for useful discussions.
This material is based upon work supported by the North Atlantic Treaty 
Organization under a Grant awarded in 1994.
This work was supported in part by 
DOE grant DE-FG02-94ER40823.

\newpage

\centerline{FIGURE CAPTIONS}
\bigskip

\begin{enumerate}

\item \label{fig1}
Normalized likelihood distribution for each of \he4 and \li7, shown as a 
function of $\eta$.  The one-peak structure of the \he4 curve
corresponds to its monotonic increase with $\eta$, while
the two-peaks for \li7 arise from its passing through a minimum in the
theoretical calculation. Systematic uncertainties have been assumed to be
top-hat distributions.

\item \label{fig2} 
Combined likelihood for simultaneously fitting \he4 and \li7,
as a function of $\eta$. Plotted is the product of the normalized likelihood
distributions shown in figure \ref{fig1} multiplied by $2\pi \sigma_4 \sigma_7$.

\item \label{fig3}
As is Figure 1, where the systematics have been assumed to be Gaussian
distributions.

\item \label{fig4}
As in figure 2, using the likelihood distributions in figure 3.

\item \label{fig5}
As is Figure 3, where a factor of 2 depletion for \li7 has been assumed.

\item \label{fig6}
As in figure 2, using the likelihood distributions in figure 5.

\item \label{fig7}
As in figure 1, adding the likelihood for D/H as given by
the QSOALS observations \cite{quas1}.

\item \label{fig8}
As in figure 2, using the likelihood distribution in figure 7.

\item \label{fig9}
As in figure 5, assuming the observed Pop II \li7 to be depleted by
a factor of 2, and including the D/H as observed in QSOALS.

\item \label{fig10}
As in figure 6, using the likelihood distribution of figure 9.

\end{enumerate}


\begin{thebibliography}{99}

\bibitem{wssok} T.P. Walker, G. Steigman, D.N. Schramm, 
 K.A. Olive and K. Kang, Ap. J. {\bf 376}
 (1991) 51.

\bibitem{he4yields} B.D. Fields, S. Dodelson, and M.S. Turner, 
Phys. Rev. {\bf D} (1993); D. Seckel, Phys. Rev. {\bf D}
 (1993); P.J. Kernan (1993), Ph.\ D. thesis,
Ohio State.

\bibitem{kr} L.M. Krauss and P. Romanelli,  ApJ, {\bf 358} (1990) 47.

\bibitem{skm} M. Smith, L. Kawano, and R.A. Malaney, 
Ap.J. Supp., {\bf 85} (1993) 219.

\bibitem{kk1} P.J. Kernan and L.M. Krauss, 
Phys. Rev. Lett. {\bf 72} (1994) 3309.

\bibitem{kk2} L.M. Krauss and P.J. Kernan, Phys. Lett. {\bf B347} (1995) 347.

\bibitem{hata1} N. Hata, R.J. Scherrer, G. Steigman, D. Thomas, and T.P.
Walker, Ap.J., {\bf 458} (1996) 637 (astro-ph/9412087).

\bibitem{osa} K.A. Olive and G. Steigman,  Ap.J. Supp. {\bf 97} (1995) 49.

\bibitem{csta} C.J. Copi, D.N. Schramm, and M.S. Turner, 
Science, {\bf 267} (1995) 192.

\bibitem{fo} B.D. Fields, and K.A. Olive,  Phys Lett {\bf B368} (1996)
103.


\bibitem{st92} G. Steigman and M. Tosi, Ap.J. {\bf 401} (1992) 15.

\bibitem{vop} E. Vangioni-Flam, K.A. Olive, and N. Prantzos, Ap.J. {\bf 
427} (1994) 618.

\bibitem{bdf} B.D. Fields, ApJ {\bf 456} (1996), 478. 

\bibitem{ytsso} J. Yang, M.S. Turner, G. Steigman, D.N. Schramm, and
 K.A. Olive, Ap.J. {\bf 281} (1984) 493.

\bibitem{st95} G. Steigman and M. Tosi, Ap.J. {\bf 453} (1995) 173.

\bibitem{Dar} A.\ Dar, J.\ Goldberg and M.\ Rudzsky, (1995) astro-ph/9405010.

\bibitem{ddp} M.H. Pinsonneault, C.P. Deliyannis, and P. Demarque,
Ap.J. Supp. {\bf 78} (1992) 179.

\bibitem{ddk} C.P. Deliyannis, P. Demarque, and S.D. Kawaler,
Ap.J. Supp. {\bf 73} (1990) 21.

\bibitem{quas1} R.F. Carswell, M. Rauch, R.J. Weymann, A.J. Cooke, 
J.K. Webb, {\it MNRAS} {\bf 268} (1994) L1; A. Songaila, L.L. Cowie,  
C. Hogan, M. Rugers, {\it Nature} {\bf 368} (1994) 599.

\bibitem{rh} M. Rugers and C. Hogan, ApJ Lett. {\bf 259} (1996) 1.

\bibitem{rw} R. Rood, and T.W. Wilson,  Ann. Rev. Astron. Astrophys.,
 (1995) 191.

\bibitem{elba}{\it
 The Light Element Abundances, Proceedings of the ESO/EIPC Workshop},
ed. P. Crane, (Berlin:Springer,1995)

\bibitem{scostv} S.T. Scully, M. Cass\'{e}, K.A. Olive, D.N. Schramm,
J.W. Truran, and E. Vangioni-Flam, astro-ph/0508086, Ap.J. {\bf 462} (1996)
in press.

\bibitem{linetal} J.L. Linsky, et al., Ap.J. {\bf 402} (1993) 694;
J.L. Linsky, et al., Ap.J. (1995) in press.

\bibitem{fer} R. Ferlet, in {\it The Proceedings of the IInd
Rencontres du Vietnam}, ed. J. Tran Thanh Van,
 (Gif Sur Yvette:Editions Frontiers, 1996)

\bibitem{linsky} J. Linsky,  talk at Aspen Center for Physics
Workshop on Primordial Nucleosynthesis, (1995).

\bibitem{bbbrw} D.S. Balser, T.M. Bania, C.J. Brockway, R.T.
Rood, and T.L. Wilson, Ap.J. {\bf 430} (1994) 667.

\bibitem{rst} R.T. Rood, G. Steigman, and B.M. Tinsley, Ap.J. 
{\bf 207} (1976) L57.

\bibitem{it} I. Iben and J.W. Truran, Ap. J. {\bf 220} (1978) 980.

\bibitem{rood} Rood, R.T., Bania, T.M., \& Wilson, T.L.
 Nature {\bf 355} (1992) 618;
Rood, R.T., Bania, T.M.,  Wilson, T.L., \& Bania, D.S. 1995,
in {\it
 the Light Element Abundances, Proceedings of the ESO/EIPC Workshop},
ed. P. Crane, (Berlin:Springer), p. 201

\bibitem{orstv} K.A. Olive, R.T. Rood, D.N. Schramm, J.W. Truran, and E.
Vangioni-Flam, Ap.J. {\bf 444} (1995) 680.

\bibitem{galli} D. Galli, F. Palla, F. Ferrini, and U. Penco,
Ap.J. {\bf 443} (1995) 536.

\bibitem{dst} M. Tosi, G. Steigman, and D.S.P. Dearborn, in {\it The
Light Element Abundances, Proceedings of the ESO/EIPC Workshop},
ed. P. Crane, (Berlin:Springer, 1995) 228.

\bibitem{vw} E. Vassiladis and P.R. Wood, Ap.J. {\bf 413} (1993) 641.

\bibitem{wwd} A. Weiss, J. Wagenhuber, and P. Denissenkov, astro-ph/9512120
(1995).

\bibitem{hogan} C. Hogan, Ap.J. Lett. (1995).

\bibitem{ch} C. Charbonnel, A. A. {\bf 282} (1994) 811.

\bibitem{wbs} G.J. Wasserburg, A.I. Boothroyd, and I.-J. Sackman,
Ap.J. {\bf 447} (1995) L37.

\bibitem{sb} I.-J. Sackman and A.I. Boothroyd, astro-ph/9512122
(1995).

\bibitem{bm} A.I. Boothroyd and R.A. Malaney,  astro-ph/9512133
(1995).

\bibitem{osst} K.A. Olive, D. N. Schramm, S.T. Scully, and J.W. Truran,
in preparation.

\bibitem{p} B.E.J. Pagel, E.A. Simonson, R.J. Terlevich and M. Edmunds, 
MNRAS {\bf 255} (1992) 325.

\bibitem{evan} E. Skillman et al., Ap.J. Lett. (in preparation) 1995.

\bibitem{iz} Y.I. Izatov, T.X. Thuan, and V.A. Lipovetsky,
Ap.J. 435 {\bf 435} (1994) 647.

\bibitem{osc} K.A. Olive, and S.T. Scully, IJMPA {\bf 11} (1996) 409.

\bibitem{skill} E. Skillman, R.J. Terlevich, R.C. Kennicutt, D.R.Garnett,
and E. Terlevich, Ap.J. {\bf 431} (1994) 172.

\bibitem{mol} P. Molaro, F. Primas, and P. Bonifacio, A.A. {\bf 295 }
(1995) L47.

\bibitem{sfosw}G. Steigman, B. Fields, K.A. Olive, D.N. Schramm, and 
T.P.  Walker, Ap.J. {\bf 415} (1993) L35.

\bibitem{wssof} T.P. Walker, G. Steigman, D.N. Schramm, K.A. Olive
and B. Fields, Ap.J. {\bf 413} (1993) 562.

\bibitem{fos} B.D. Fields, K.A. Olive, and D.N. Schramm, Ap.J. {\bf 435}
(1994) 185.

\bibitem{vcfo} E. Vangioni-Flam, M. Cass\'e, B.D. Fields, and K.A. Olive,
Ap.J. (1996) in press.

\bibitem{hata2} N. Hata, R. J. Scherrer, G. Steigman, D. Thomas,
T. P. Walker, S. Bludman and P. Langacker,
Phys.\ Rev.\ Lett.\ {\bf 75} (1995) 3977.

\bibitem{osb} K.A. Olive and G. Steigman, Phys. Lett. {\bf B354 } (1995) 357.

\bibitem{ossty}   K.A. Olive, D.N. Schramm, G. Steigman, M.S. Turner, and J.
Yang,  Ap.J. {\bf 246} (1981) 557.

\bibitem{fko1} B.D.\ Fields, K.\ Kainulainen and K.A.\ Olive, 
 hep-ph/9512321 (1995).

\bibitem{cstb} C.J. Copi, D.N. Schramm, and M.S. Turner,  Ap.J. {\bf 455} (1995)
L95; C.J. Copi, D.N. Schramm, and M.S. Turner,  
Phys.\ Rev.\ Lett., (1995)
submitted.

\bibitem{Sarkar} S. Sarkar, OUTP-95-16P, hep-ph/9602260 (1996).




\end{thebibliography}
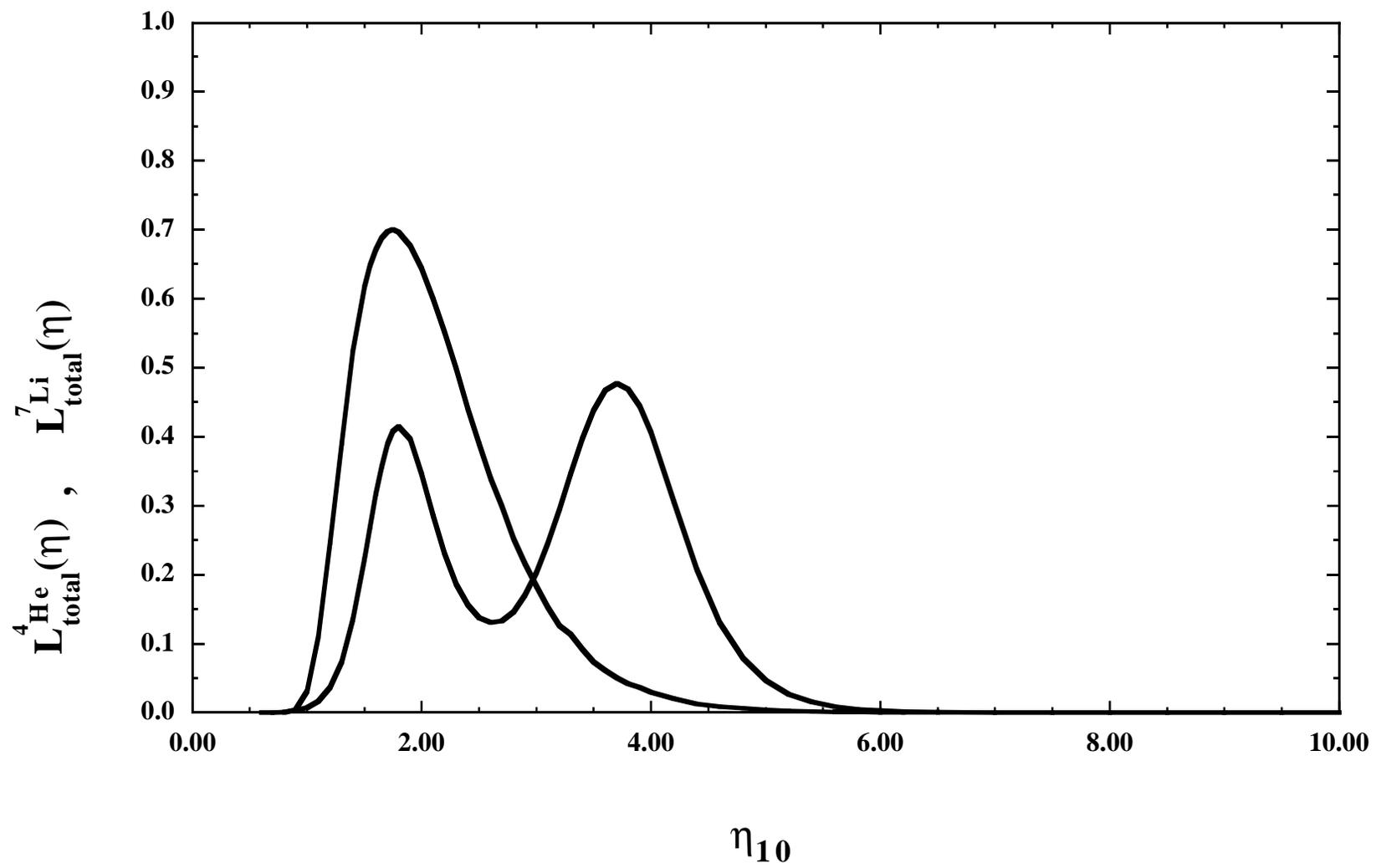
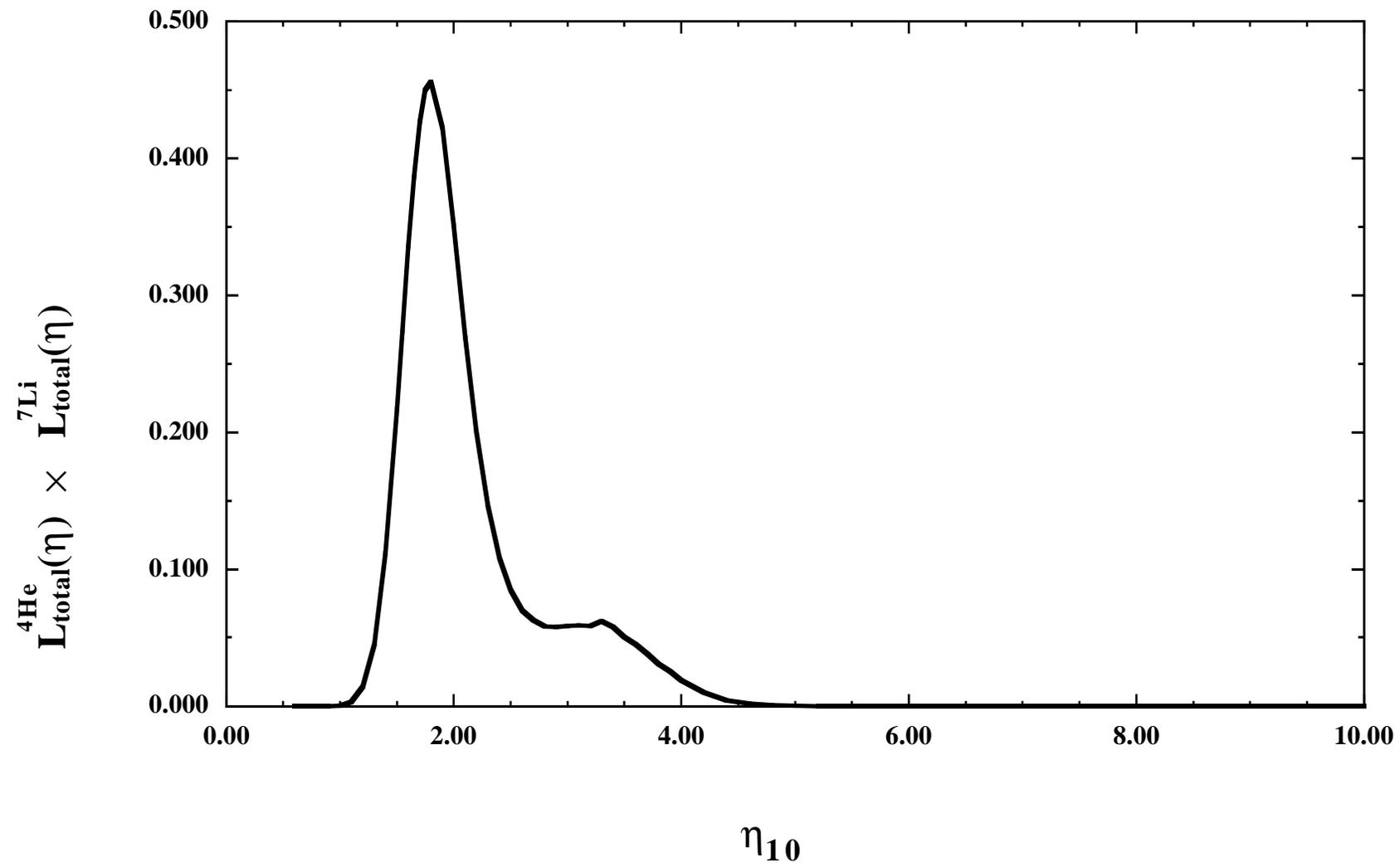
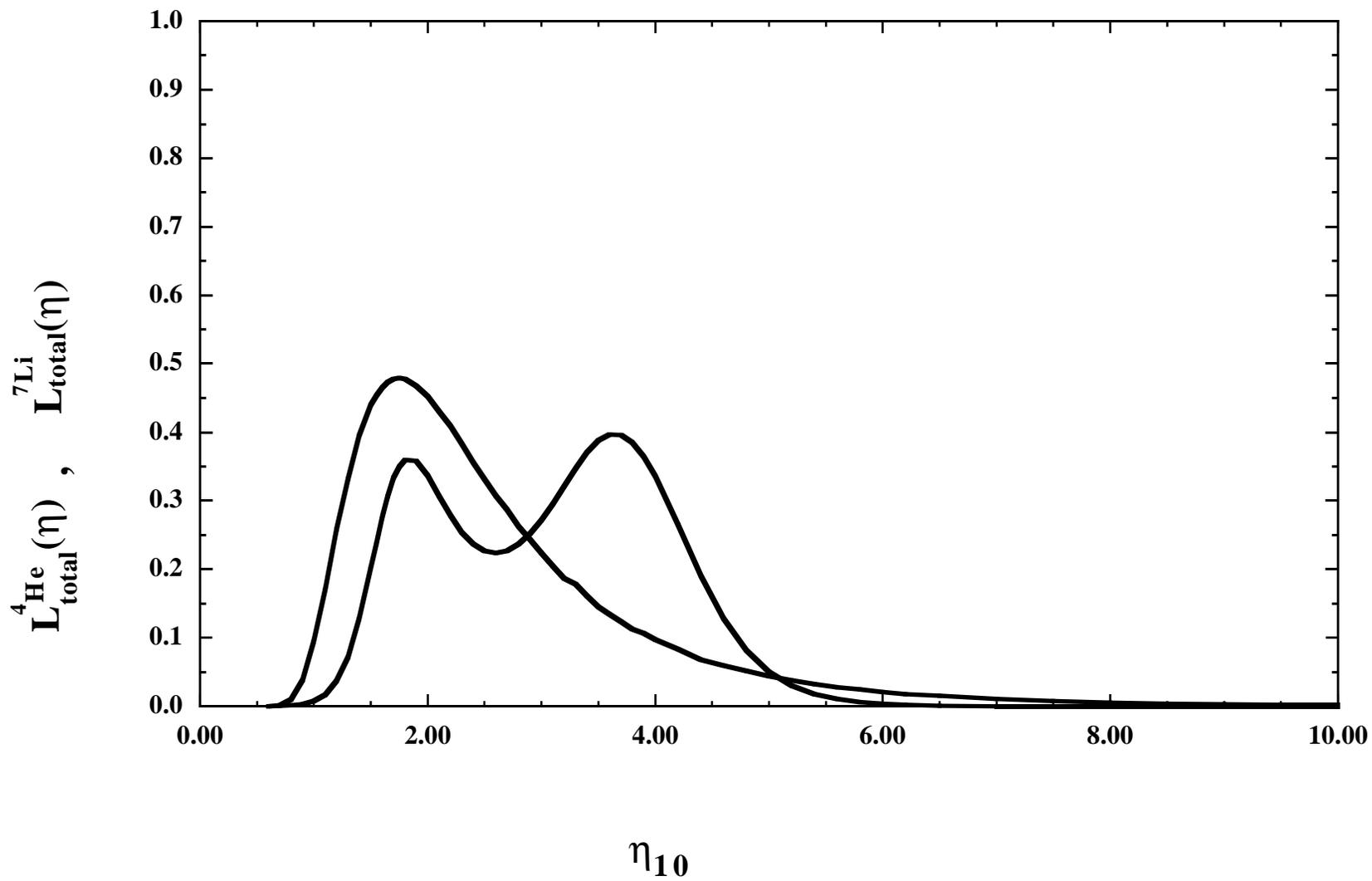
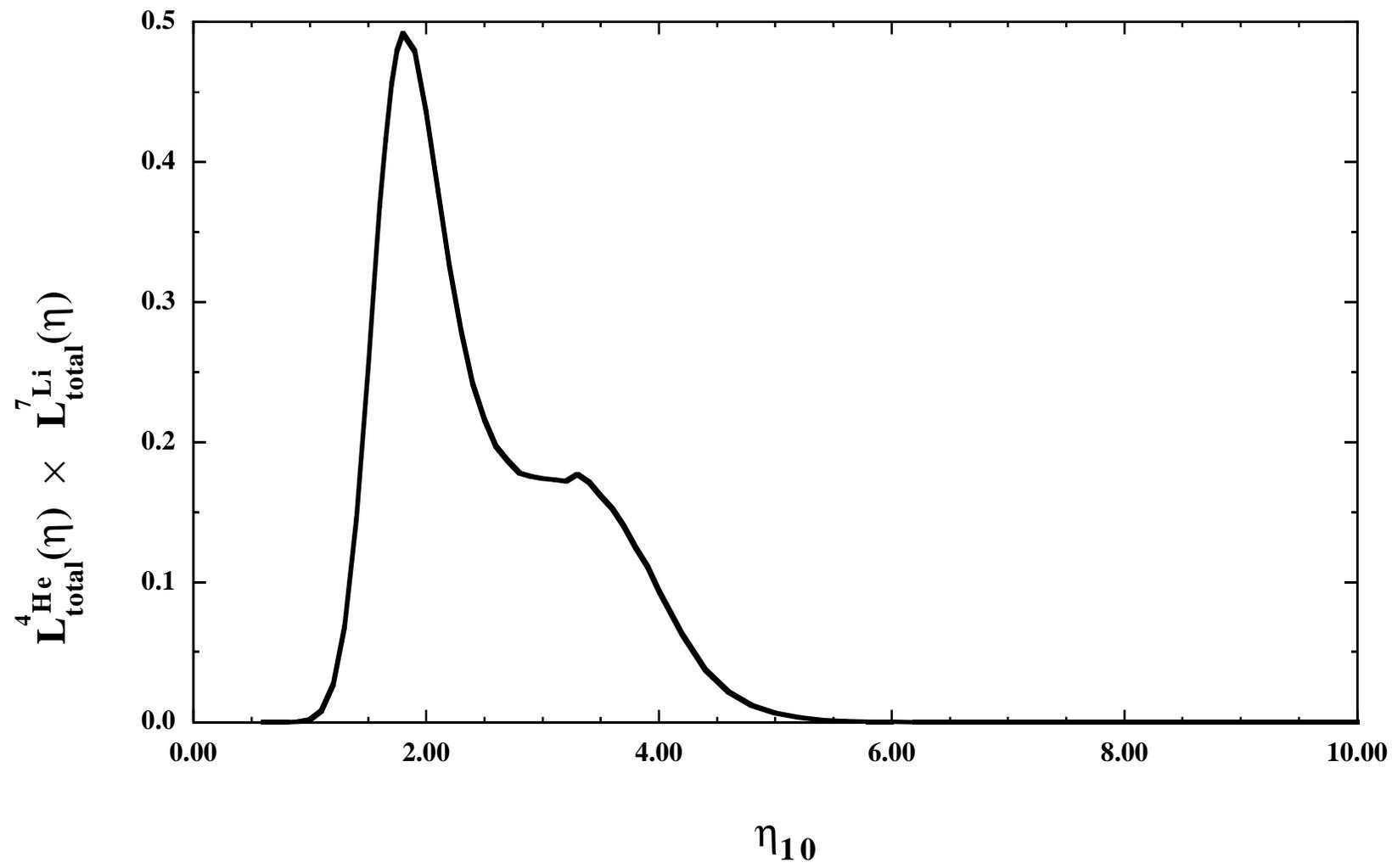
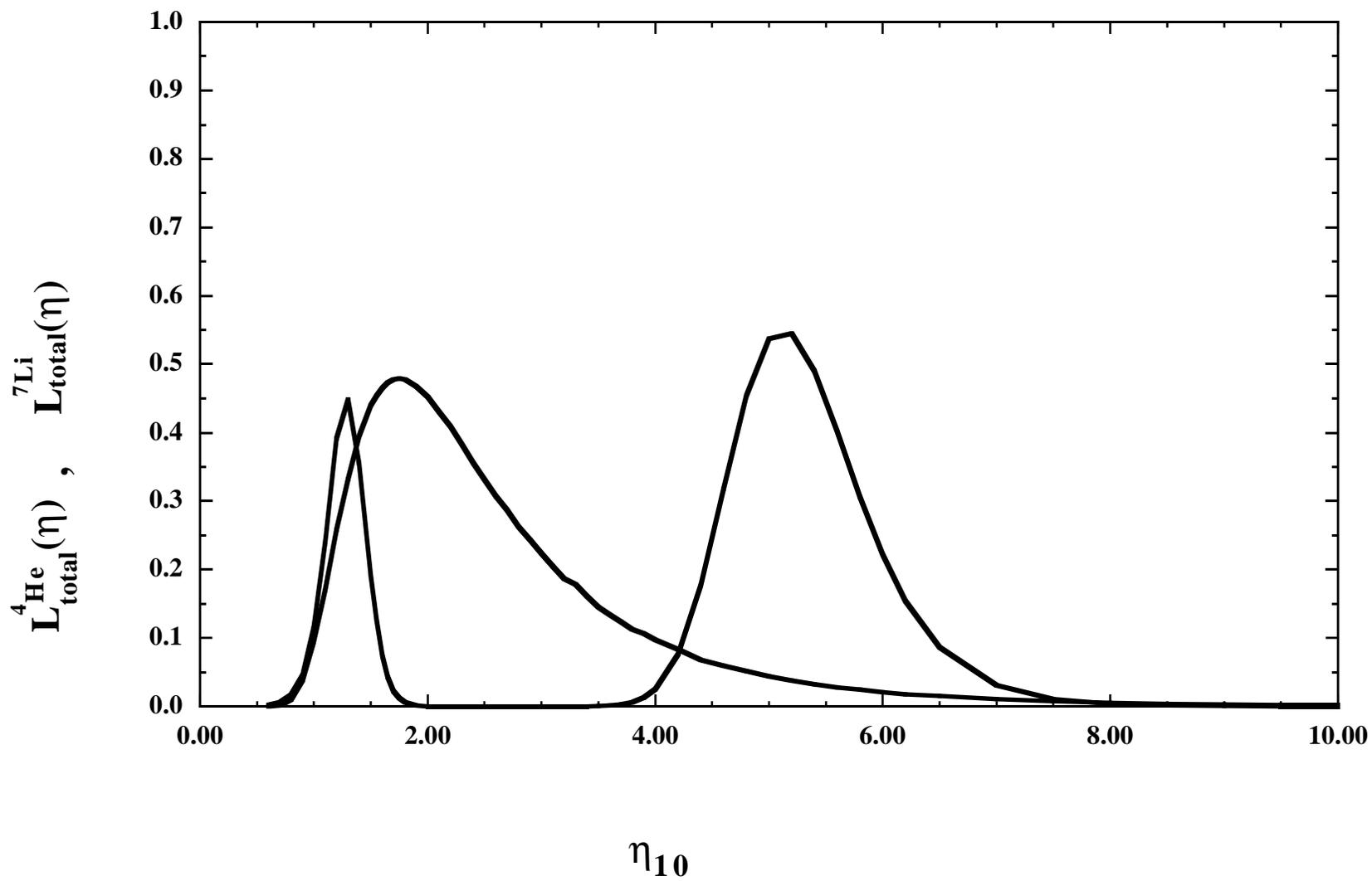
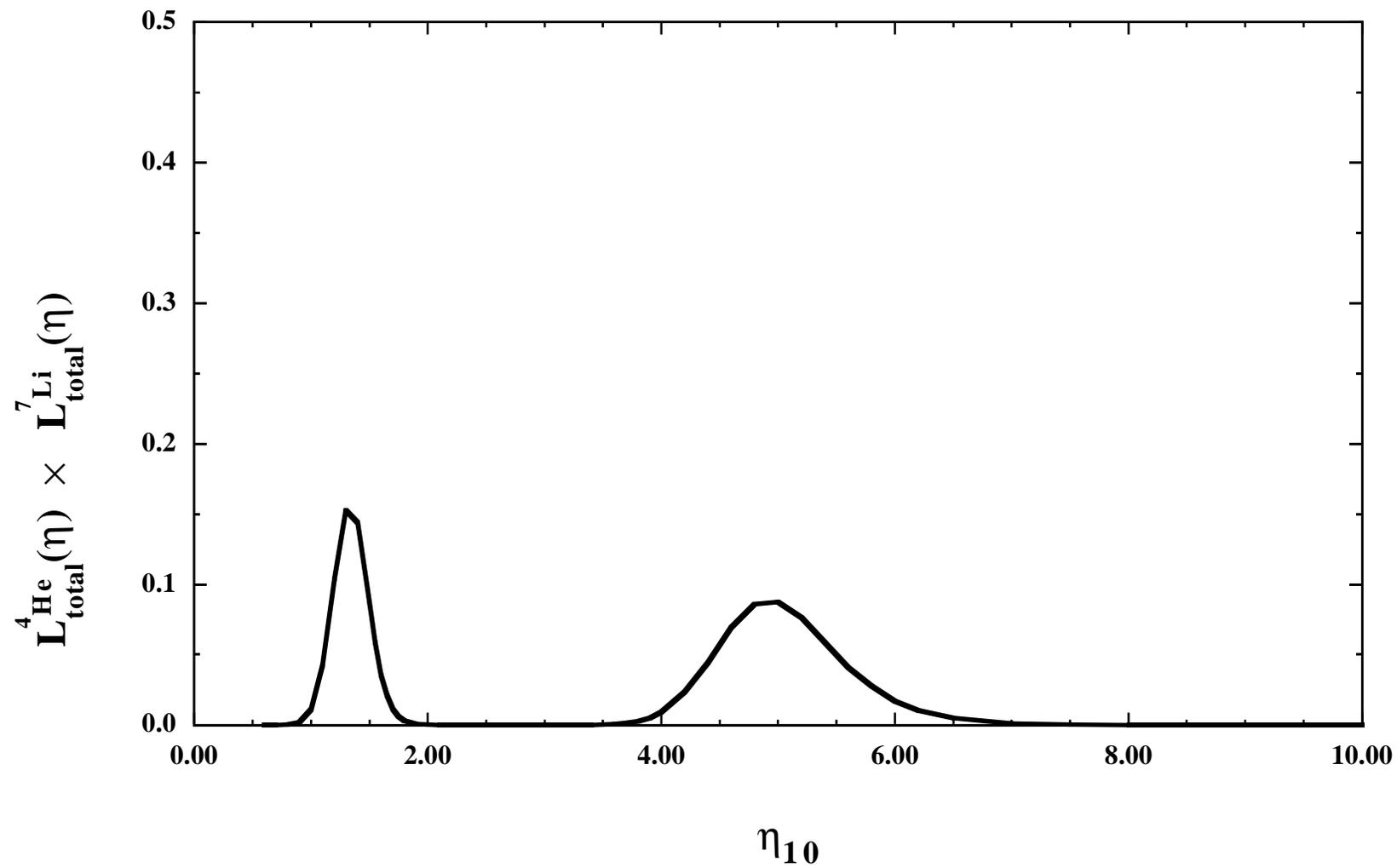
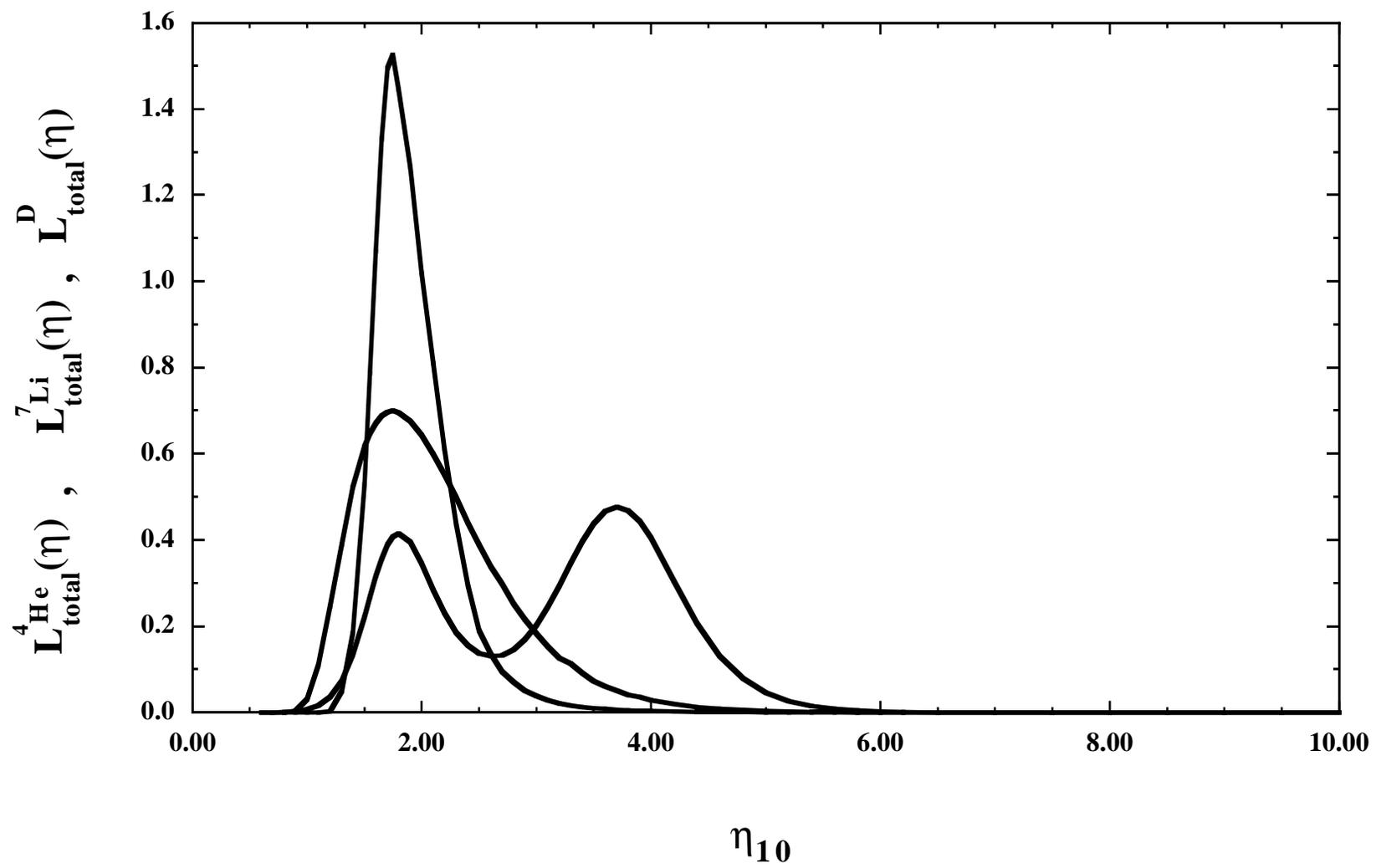
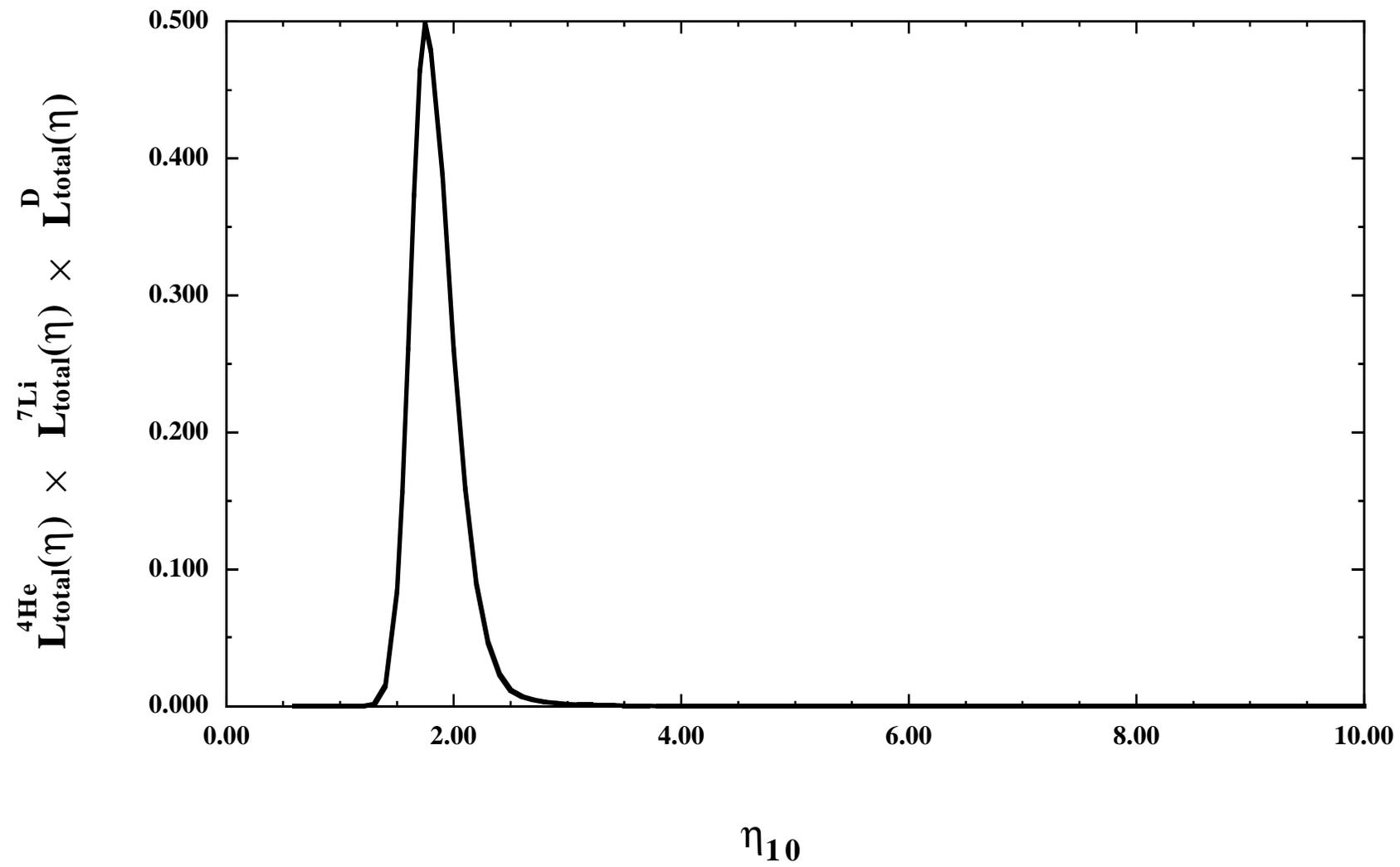
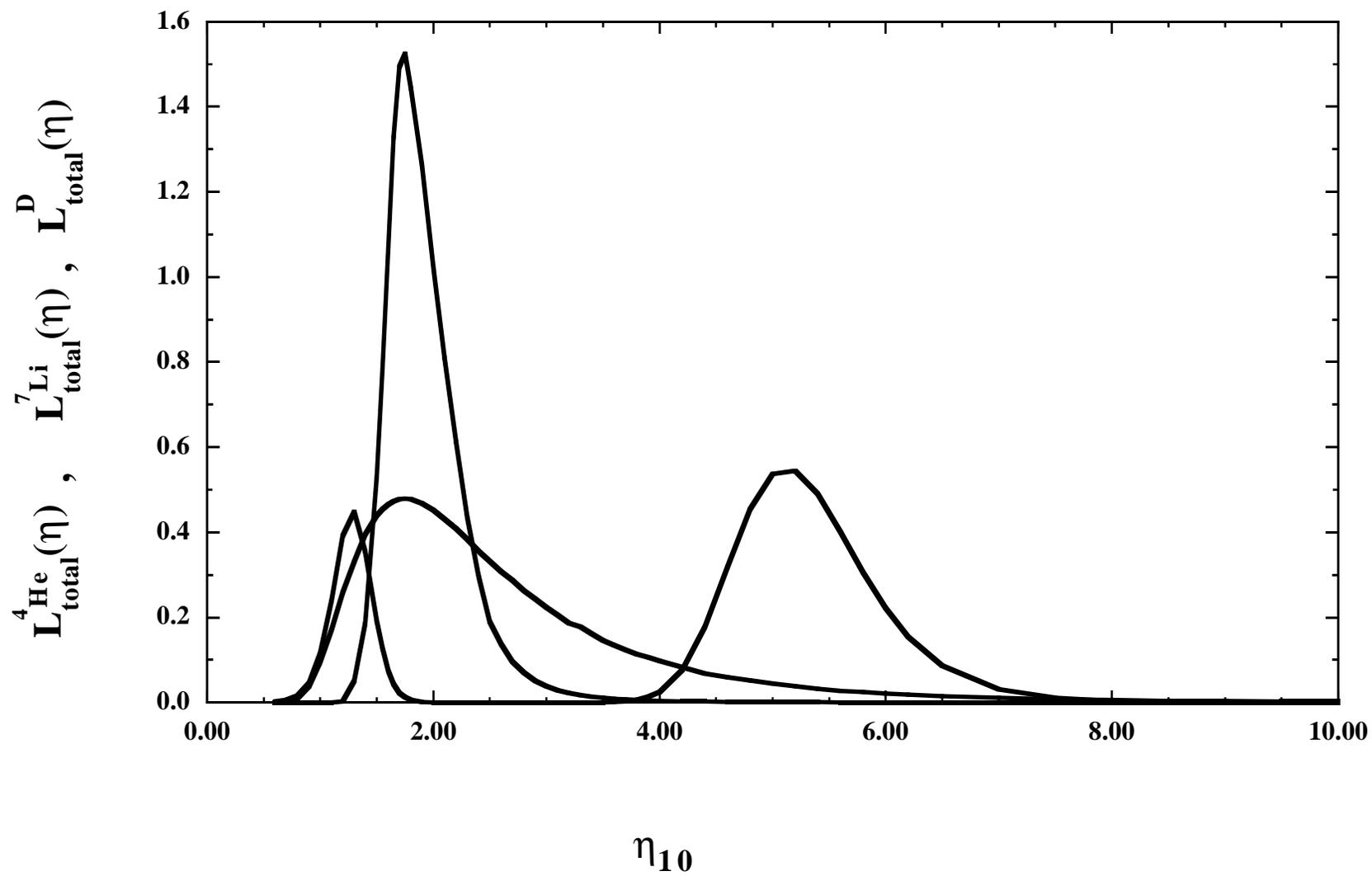
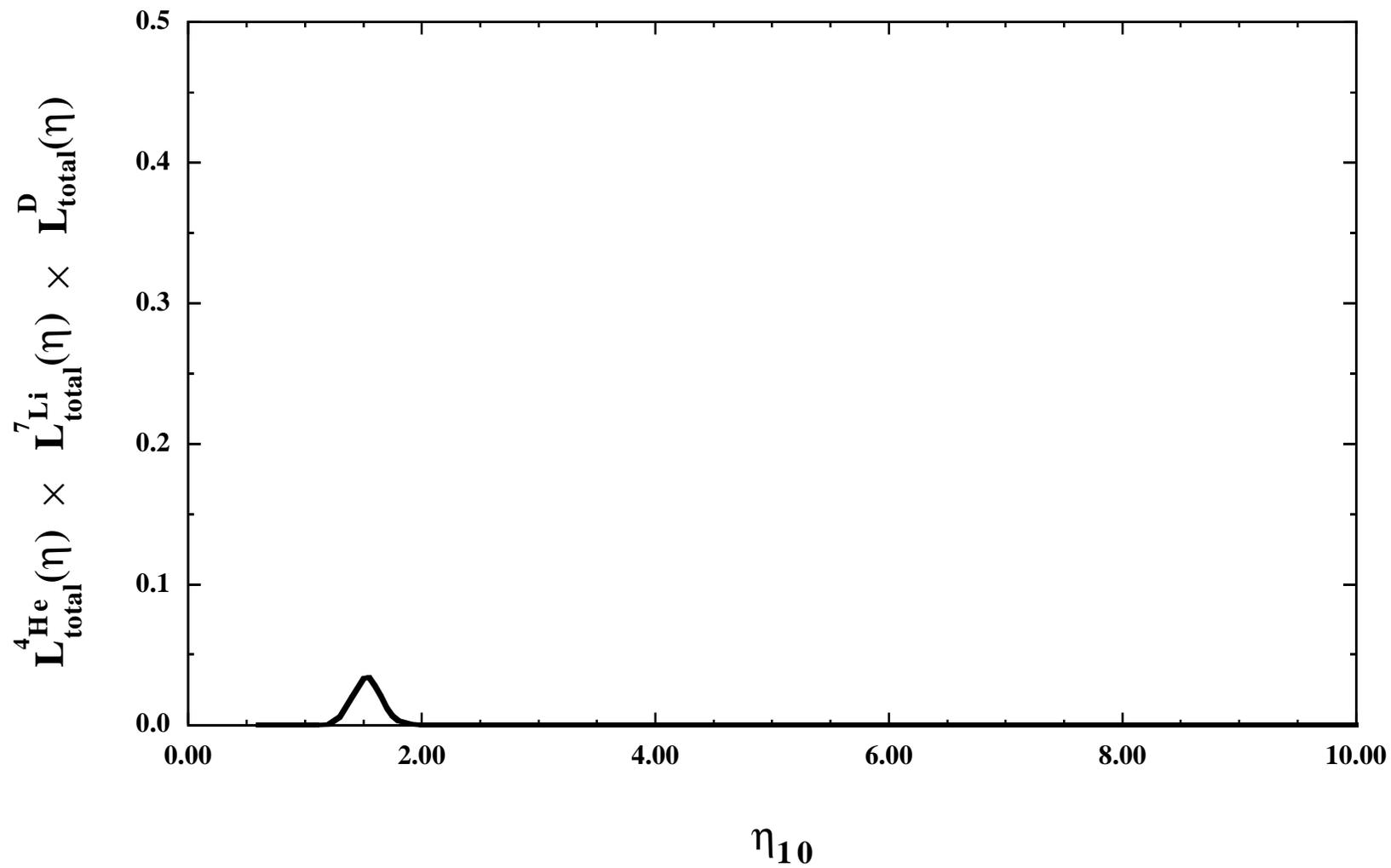
\end{document}